\documentclass[prb,aps,showpacs,twocolumn,preprintnumbers,amsmath,amssymb,superscriptaddress]{revtex4-1}
\usepackage{amsmath,amssymb,amsfonts}
\usepackage{graphicx}
\usepackage[colorlinks=True,linkcolor=red,citecolor=blue,urlcolor=blue]{hyperref}
\usepackage{xcolor}
\usepackage{chngcntr}
\usepackage{braket}
\usepackage{hyperref}
\usepackage{nicefrac}
\usepackage{bm}
\usepackage{multirow}
\usepackage{mathtools}

\usepackage{ulem}

\begin{document}
\bibliographystyle{apsrev4-1}
\title{Intrinsic efficiency of injection photocurrents in magnetic materials} 

\author{Alonso J. Campos-Hernandez}
\affiliation{IMDEA Nanociencia, 28049 Madrid, Spain}

\author{Yang Zhang}
\affiliation{Department of Physics, Massachusetts Institute of Technology, Cambridge, MA 02139, USA}

\author{Adolfo G. Grushin}
\affiliation{Univ. Grenoble Alpes, CNRS, Grenoble INP, Institut N\'eel, 38000 Grenoble, France}

\author{Fernando de Juan}
\affiliation{Donostia International Physics Center, P. Manuel de Lardizabal 4, 20018 Donostia-San Sebastian, Spain}
\affiliation{IKERBASQUE, Basque Foundation for Science, Maria Diaz de Haro 3, 48013 Bilbao, Spain}

\date{\today}
\begin{abstract}
The generation of shift and ballistic photocurrents in non-centrosymmetric materials represents a promising alternative mechanism for light energy harvesting. Many studies have focused on finding the best suited materials by maximizing the photocurrent magnitude, but estimating the actual efficiency  requires knowledge of the light-induced DC photoconductivity and is rarely considered. Using the recently proposed jerk current as photoconductivity, in this work we show that only ballistic photocurrents have finite efficiency in the limit of large relaxation times $\tau$. Moreover, at zero temperature the only ballistic current which is finite for unpolarized light is the magnetic injection current, only present in magnetic materials. We present a band structure expression for the efficiency of such photocurrent, showing  that it scales as $(\hbar \Omega-E_g)^2$ near the band edge, and we present its frequency dependence for a simple tight-binding model. Our work provides a new tool to guide the search for efficient energy harvesting based on the magnetic injection current. 
\end{abstract}
\maketitle

\section{Introduction}
\label{sec:intro}

The bulk photovoltaic (or photogalvanic) effect, which enables photocurrent generation in crystalline materials that break inversion symmetry~\cite{BS80}, constitutes a promising alternative mechanism for energy harvesting applications. While the phenomenon itself has been known for a long time, a renewed interest in the problem is on the rise thanks to both novel theoretical approaches~\cite{Young12,Tan16,Cook17,Fregoso17,Wang17,Ibanez18,Tan19} to guide the search for an enhanced response and promising new material candidates~\cite{Spanier16,Burch17,Zhang19,Li21,Akamatsu21}. 

While the generation of the largest possible photocurrent $I$ is an interesting problem in itself, it is well known that to maximize the efficiency for light energy harvesting the full bias dependence $I(V)$ needs to be known. This requires knowledge of the conductivity of the photogenerated carriers in the presence of applied light, known as the photoconductivity. Therefore, the problem of realistically modeling  the bulk photovoltaic efficiency is of great interest~\cite{Tan16,MorimotoShift18}, but also highly complex, as there are no apparent simple guiding principles. Rather than attempting realistic predictions, the aim of this work is to ascertain whether there is any situation, simplified as it might be, where the efficiency of a photocurrent solar panel can be computed as an intrinsic band-structure property of a given material. Finding these instances will be helpful for an order-of-magnitude estimation of efficiencies, will serve to guide the search for useful materials and symmetry groups, and might even be helpful for high-throughput ab-initio calculations. 

The bulk photovoltaic effect, also known as photogalvanic effect (PGE), can be subdivided into two classes known as the linear (LPGE) and circular (CPGE) photogalvanic effects. The CPGE switches sign with the circular component of the light polarization, and therefore averages to zero for unpolarized light. The LPGE, however, can  be finite for unpolarized light in the crystal classes with a polar axis~\cite{BS80} which sets a preferred direction for current flow. This property singles out polar crystals as the only viable option for energy harvesting via LPGE. 

Microscopically, there are two main mechanisms that contribute to the photogalvanic effect in semiconductors, known as the shift and ballistic currents~\cite{BS80,Sturman20}. The shift current is an intrinsic band structure effect. In contrast, the ballistic current generically involves scattering for example from disorder or phonons and requires knowledge of extrinsic properties like the carrier lifetime $\tau$.  At zero temperature and in the relaxation time approximation, however, an expression for the ballistic current can be derived which is also an intrinsic band structure property multiplied by a prefactor of $\tau$, which is referred to as the injection current~\cite{SipeShkrebtii,AversaSipe,NastosSipe}. In the cleanest samples, the injection current always dominates over the shift current since $\tau$ can be made arbitrarily large. However, in the presence of time-reversal symmetry $\mathcal{T}$ (i.e. for non-magnetic materials), the shift current only gives rise to LPGE, while the injection current only gives rise to CPGE, so there is no injection current for unpolarized light. The phonon-induced ballistic current does give rise to an LPGE linear in $\tau$ \cite{Dai21} even with time-reversal symmetry, but the effect requires thermal population of phonons and vanishes at zero temperature. For magnetic materials, however, the linear in $\tau$ injection current does have an extra LPGE component~\cite{DFG20,ZhangCrI3} which has been the subject of many recent studies~\cite{ZhangCrI3,Fei20,Wang20,Ahn20,Holder20,Watanabe21,Watanabe21b,Merte21,Okumura21}. This \textit{magnetic injection} LPGE is therefore a promising candidate for energy harvesting. 

In this work we will derive a simple estimate for the efficiency of photogalvanic effects in the relaxation time approximation at zero temperature. Using the recently derived jerk current~\cite{Jerk18,Fregoso19,Ventura20} as an estimate for the photoconductivity, which is also a band-structure property scaling with a $\tau^2$ prefactor, it will be shown that the shift current efficiency vanishes as $1/\tau^2$ as $\tau \rightarrow \infty$, while the magnetic injection current efficiency is $\tau$-independent and is therefore an intrinsic band-structure property. Within our simplified picture, this reveals magnetic injection LPGE as the only viable mechanism to generate finite light harvesting efficiency in clean, homogeneous systems at zero temperature. In the rest of this work, we will present the efficiency derivation and the explicit expressions, and evaluate them for suitable symmetry-constrained two band models for magnetic systems to illustrate the method and provide numerical values for the efficiency in terms of band parameters. 

\section{Photogalvanic effects and efficiency}

\begin{figure*}
    \centering
    \includegraphics[width=0.9\textwidth]{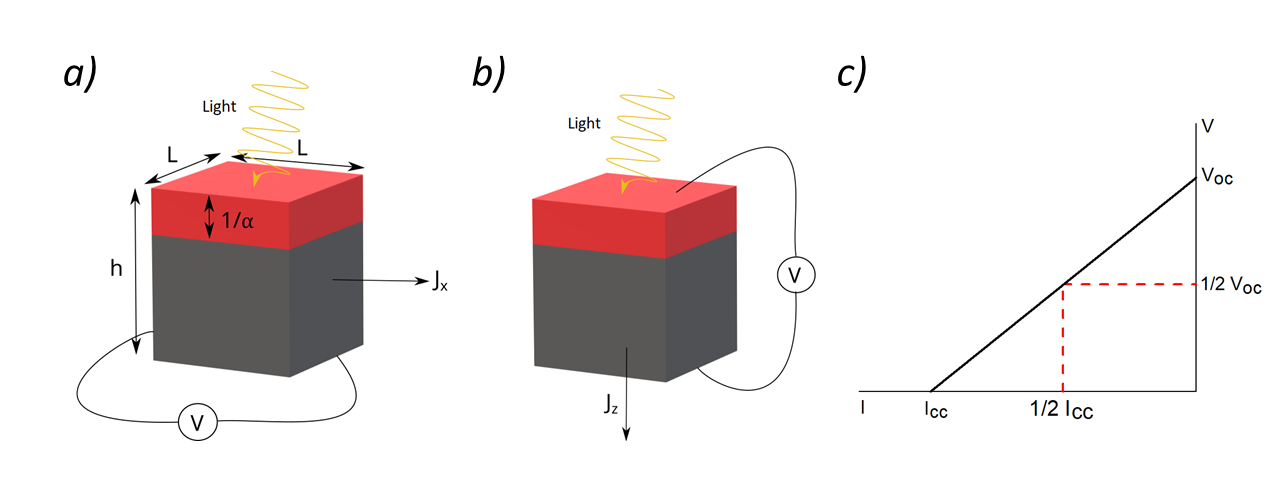}
    \caption{Schematic photodevice considered in this work and its ideal I-V curve. A sample of surface area $L^2$ and thickness $h$ is irradiated by light, which is absorbed within an attenuation length $1/\alpha$. a) and b) show the two possible current geometries considered in the text. Subfigure c) shows the relation between the device's ideal V and I. The red dashed lines indicate the values maximizing energy harvesting efficiency.}
    \label{fig:Device}
\end{figure*}

We consider the efficiency for power generation by a monochromatic light beam of frequency $\Omega$. For ease of notation we will not write the $\Omega$ dependence of optical quantities, which should be understood. To compute the efficiency we adapt the calculation presented in Refs.  \cite{Tan16,MorimotoShift18}. The LPGE response to light is described phenomenologically by 
\begin{align}
J^{(2)}_i = \sigma_{ijk}^{(2)}(E_jE_k^*+E_kE_j^*),    
\end{align}
where $\sigma_{ijk}^{(2)}$ is the photoconductivity, $J^{(2)}_i$ is the current density and the electric field is $E_j = \tilde{E}_j \mathcal{E}$ with $\tilde{E}_j$ the polarization vector and $\mathcal{E}$ the magnitude of the electric field.   

Defining the light intensity $\mathcal{I} = c\epsilon_0 \mathcal{E}^2/2$, we can define a responsivity
\begin{align}
J^{(2)}_i = \kappa_i \mathcal{I},    
\end{align}
where $\kappa_i = \tfrac{4}{c\epsilon_0} \sigma_{ijk}^{(2)}\tilde{E}_j\tilde{E}_k$ for a given polarization $\tilde{E}_j$. To see how this photocurrent can be used to harvest energy from light, we consider a device of thickness $h$ and lateral dimensions $L$ as shown in Fig. \ref{fig:Device}. In a bulk sample, the current does not actually respond to the full initial intensity due to absorption effects. When light is irradiated, part of it is absorbed by the material and the intensity decays exponentially with distance from the surface as $\mathcal{I}(z) = \mathcal{I}_0 e^{-\alpha |z|}$ with $\alpha$ the attenuation coefficient. The average intensity on the whole sample is therefore 
\begin{align}
\mathcal{\bar{I}} = \frac{1}{h} \int_0^h dz\mathcal{I}(z) = \frac{\mathcal{I}_0}{\alpha h}(1-e^{-\alpha h}) \approx \frac{\mathcal{I}_0}{\alpha h},
\end{align}
where in the last approximation we used that for a bulk sample $h \gg 1/\alpha$. The total generated current density is therefore expressed as
\begin{align}
\label{eq:J2cc}
J_i^{(2)} = \kappa_i \mathcal{\bar{I}} = \frac{\kappa_i}{\alpha h}\mathcal{I}_0,
\end{align}
where $\tfrac{\kappa_i}{\alpha }$ is known as the Glass coefficient. 

The current density $J_i^{(2)}$ determines the closed circuit current $I_{\rm cc} = J^{(2)}_i A$ for a given cross section $A$, and is the current that would be produced if the device was connected to a resistor with $R=0$ and no voltage drop, as shown in Fig \ref{fig:Device}. In reality, the device will be connected to a finite resistance, and to know the voltage drop we need to know the photoconductivity, which is the DC conductivity of the sample in the presence of light irradiation
\begin{align}
\label{eq:DCphotocond}
J_i^{(3)} = \sigma_{ijkl}^{(3)}(E_jE_k^*+E_kE_j^*) E^{0}_l ,   
\end{align}
where $E^0_l$ is the electric field corresponding to the applied voltage. We can again define the analog response to light intensity as 
\begin{align}
\label{eq:J3}
J_i^{(3)} = \Sigma_{ij} \frac{\mathcal{I}_0}{\alpha h} E^{0}_j,    
\end{align}
with $\Sigma_{il} = \tfrac{4}{c\epsilon_0} \sigma_{ijkl}^{(3)}\tilde{E}_j\tilde{E}_k$. 

The maximum voltage drop occurs when the device is not connected to a resistance and the photocurrent is fully compensated by the current induced by the voltage drop
\begin{align}
J_i^{(2)} + J_i^{(3)} = (\kappa_i - \Sigma_{ij}  E^{0}_j)   \frac{\mathcal{I}_0}{\alpha h} = 0,
\end{align}
which defines the open circuit electric field $E^0_i = \kappa_i/\Sigma_{ii}$ for a given direction $i$, and the corresponding open circuit voltage $V_{\rm oc} = E^0_i d$ where $d$ is the distance across which the voltage drop occurs.

When the device is operated at a finite resistance $R$, the total power that can be extracted is $P(R) = I(R)V(R)$. The power of the electromagnetic field of light through a cross section $L^2$ is $W=\mathcal{I}_0 L^2$, and the efficiency $\eta$ is the ratio of the power extracted to light power, $\eta = P/W$. Since $I(V)$ is linear in $V$ for the LPGE, the product $IV$ is maximized when $I = \tfrac{1}{2}I_{\rm cc}$ and $V = \tfrac{1}{2}V_{\rm oc}$ so $P = I(R)V(R) = \tfrac{1}{4} I_{\rm cc}V_{\rm oc}$, as shown in Fig. \ref{fig:Device}c).  Substituting we have 
\begin{align}
\eta = \frac{P}{\mathcal{I}_0 L^2} =   \frac{1}{4}\frac{(J^{(2)}_i A) (E^0_i d)}{\mathcal{I}_0 L^2} = \frac{1}{4} \frac{\kappa_i^2}{\Sigma_{ii} \alpha h} \frac{Ad}{ L^2}.
\end{align}
For both current directions considered in Fig. \ref{fig:Device} we have $A d = L^2 h$ so 
\begin{align}
\eta = \frac{1}{4} \frac{\kappa_i^2}{\Sigma_{ii} \alpha} .\label{eq:eff}
\end{align}
As anticipated, since for injection currents $\kappa_i \propto \tau$ while $\Sigma_{ii}\propto\tau^2$, then the injection current efficiency is $\tau$ independent. For shift currents, where $\kappa_i$ is independent of $\tau$, the efficiency vanishes as $1/\tau^2$ as $\tau \to \infty$. Because of this, in this work we only consider injection currents. 

\section{Microscopic mechanisms}

The previous expression for the efficiency is written in terms of the generic photogalvanic current $\kappa_i$, photoconductivity $\Sigma_{ij}$ and attenuation constant $\alpha$. As discussed in the previous sections, we now assume a specific microscopic mechanism for the photogalvanic effect (the injection current), and for the photoconductivity (the jerk current), which are the dominant mechanisms when $\tau \to \infty$. As shown in Appendix \ref{AppOptics}, these two quantities, along with the attenuation constant, can be computed for a generic Bloch Hamiltonian $H(k)$ in the relaxation time approximation with the expressions
\begin{widetext}
\begin{align}
    \alpha &= \frac{ \sigma_{ij}^{(1)}}{c \epsilon_0} = - \frac{1}{c \epsilon_0} e^2 \pi \Omega \int \frac{d^3k}{(2\pi)^3} \sum_{n \neq m} f_{mn} {\rm Re}[r^i_{nm}r^j_{mn}] \delta(\hbar \Omega - \epsilon_{nm}) \label{Eq:absorption},\\
\kappa_i &= \frac{4}{c\epsilon_0} \sigma_{ijk}^{(2)} = - \tau \frac{4}{c\epsilon_0}   \dfrac{ \pi e^3}{2\hbar^2} \int \frac{d^3k}{(2\pi)^3} \sum_{n \neq m} f_{nm}\partial_i\epsilon_{mn} {\rm Re}[r^j_{nm}r^k_{mn}]\delta(\hbar \Omega - \epsilon_{nm})\label{Eq:maginj},\\
\Sigma_{ij} &= \frac{4}{c\epsilon_0} \sigma_{ijkl}^{(3)} = - \tau^2 \frac{4}{c\epsilon_0}  \frac{\pi e^4}{18 \hbar^3} \int \frac{d^3k}{(2\pi)^3}\sum_{n\neq m} f_{mn} \partial_{i} \partial_l \epsilon_{nm}{\rm Re}[r^j_{nm}r^k_{mn}] \delta(\hbar \Omega - \epsilon_{nm}).\label{Eq:jerk}
\end{align}
\end{widetext}
In these equations $f_n$ is the Fermi function which we take at zero temperature, $f_{nm} = f_n - f_m$, the Bloch eigenstates are $H \left|n\right> = \epsilon_n  \left|n\right>$, $\epsilon_{nm} = \epsilon_n -\epsilon_m$, and $r_{nm}^a = i \left<n|\partial_{k_a}m\right>$ for $n\neq m$ and zero otherwise is the interband position matrix element. 

The similarity of these three expressions allows us to write the efficiency in a compact form by using the symbolic notation for the integration $\int_k = \int \frac{d^3k}{(2\pi)^3}\sum_{n\neq m} f_{mn}\delta(\hbar \Omega - \epsilon_{nm})$. Taking for simplicity all indices in the $x$ direction we find 
\begin{align}
\label{eq:eta}
\eta &= \frac{1}{4} \frac{\kappa_x^2}{\Sigma_{xx} \alpha} =   \frac{9}{ 2\hbar\Omega}\frac{[\int_k \partial_{x}\epsilon_{nm}r^x_{nm}r^x_{mn}]^2}{[\int_k \partial_{x}^2\epsilon_{nm}r^x_{nm}r^x_{mn}] [\int_k r^x_{nm}r^x_{mn}]},
\end{align}
which explicitly reveals its dimensionless character and its independence of $\tau$.

\section{Band edge and tight binding models}

To illustrate the calculation of the efficiency, we consider the simplest model for a semiconductor, a generic two band model of the form
\begin{equation}
\label{eq:H2band}
H = f_0(k) + \sum_\alpha f_\alpha(k) \sigma_\alpha,
\end{equation}
where $\sigma_\alpha$ are the Pauli matrices with $\alpha=1,2,3$, and the functions $f_0(k)$ and $f_\alpha(k)$ may correspond to matrix elements of a tight binding model, or to a power expansion in $k$ like the one obtained in the $k.p$ approximation. The computation of the relevant integrals is particularly simple for a two-band model. The energy dispersion is 
\begin{align}
\epsilon_{1} &= f_0(k) + \epsilon,\\
\epsilon_{2} &= f_0(k) - \epsilon,
\end{align}
where $\epsilon = \sqrt{\vec f ^2}$ and the optical matrix element can be obtained from the sum rule
\begin{align}
r^a_{12} r^b_{21}+r_{12}^b r^a_{21} = - \frac{v_{11}^{ba} -\partial_{k_b}\partial_{k_a}\epsilon_1}{\epsilon_{12}},
\end{align}
which is derived in Appendix \ref{App:Identities}. Here note that neither $\epsilon_{12}$ or $v_{11}^{bc} -\partial_{k_b}\partial_{k_c}E_1$ depend on $f_0$, so we can set $f_0=0$ for the rest of this work.

The explicit solution of the two band model in Appendix \ref{App:TwoBand} gives
\begin{equation}
r^a_{12} r^b_{21}+r_{12}^b r^a_{21} = \frac{   \partial_a f_\alpha \partial_b f_\alpha}{2\epsilon^2} - \frac{f_\alpha \partial_a f_\alpha f_\beta \partial_b f_\beta}{2\epsilon^4}
\end{equation}
With these ingredients, the three response functions in Eqs. \ref{Eq:absorption}-\ref{Eq:jerk} can be computed for any model of the form of \eqref{eq:H2band}. 

\subsection{Generic band edge model}

We now consider a semiconductor with a band gap $E_g = 2M$, and expand the Hamiltonian around the momentum where this minimum gap is realized, which we assume to be the $\Gamma$ point for simplicity. We expand taking as basis functions the eigenstates of $H$ at this point, so that the Hamiltonian at $k=0$ must be diagonal. This system is assumed to break inversion, time-reversal, and spin rotation symmetries. The most general Hamiltonian is given by
\begin{align}
f_1 &= v_{1,i}k_i + A_{1,ij}k_ik_j, 
\label{two-band-model-1}\\
f_2 &= v_{2,i}k_i + A_{2,ij}k_ik_j, 
\label{two-band-model-2} \\
f_3 &= M +v_{3,i}k_i +  A_{3,ij}k_ik_j,
\label{two-band-model-3}
\end{align}
and we demand that $\partial_i \epsilon_{1,2}(k)|_{k=0} =0$ so that $k=0$ represents the band edge. This sets $v_{3,i}=0$. This model generically shows the asymmetry $\epsilon_{1,2}(k) \neq \epsilon_{1,2}(-k)$ required for the magnetic injection, which occurs due to the breaking of both inversion and time-reversal symmetries. Near the band edge the leading term in the asymmetry is 
\begin{align}
\epsilon_{1}(k) - \epsilon_{1}(-k) = \frac{2}{M}(A_{1,ij}v_{1,k}+A_{2,ij}v_{2,k})k_ik_jk_k,\label{asy}
\end{align}
as shown in Appendix \ref{App:Scaling}. This model is valid only for frequencies near the band edge $\Omega = E_g +\delta \Omega$ and can be used to compute the efficiency when $\delta \Omega \rightarrow 0$. In Appendix \ref{App:Scaling} we prove that while optical absorption and jerk current both scale as $(\delta \Omega)^{1/2}$, the magnetic injection current scales as $(\delta \Omega)^{3/2}$ and according to Eq. \ref{eq:eff} the efficiency near the band edge vanishes as
\begin{align}
\eta \propto (\delta \Omega)^{2} .   
\end{align}
This result precludes the estimation of an intrinsic efficiency for the band edge, and suggests that full band models are better suited to provide such figure for the full frequency dependent efficiency. A tight binding model is considered below for this purpose.  

\subsection{Symmetry constraints}

Since our aim is to model materials with a finite respose to unpolarized light, we now consider restricting this Hamiltonian to comply with the symmetries of a polar point group, as discussed in the introduction. The polar point group with the highest symmetry is $C_{6v}$, which has the allowed LPGE components $\sigma^{(2)}_{zxx} = \sigma^{(2)}_{zyy}$, $\sigma^{(2)}_{zzz}$ and $\sigma^{(2)}_{xxz} = \sigma^{(2)}_{yyz}$, assuming the polar axis in the $z$ direction. For normal incident light the photocurrent from unpolarized light is proportional to $\sigma^{(2)}_{zxx}+\sigma^{(2)}_{zyy} =2\sigma^{(2)}_{zxx}$ and indeed flows along the polar axis $z$, reproducing the situation in Fig. \ref{fig:Device}(b). To reproduce the situation in Fig. \ref{fig:Device}(a), we consider rotating the coordinate system so that the $z$ direction, which is still the polar axis, lies in the plane, while the surface normal is the $y$ direction. In this case, the photocurrent will be proportional to $\sigma^{(2)}_{zxx}+\sigma^{(2)}_{zzz}$. While this shows that a photocurrent is generically produced from unpolarized light, for clarity in reporting our numerical results we will present $\sigma^{(2)}_{zxx}$ only, assuming linear polarization in the $x$ direction, as this component is relevant for both this direction and the $z$-incidence direction. 

\subsection{Tight binding Hamiltonian}

\begin{figure*}
    \centering
    \includegraphics[width=0.8\textwidth]{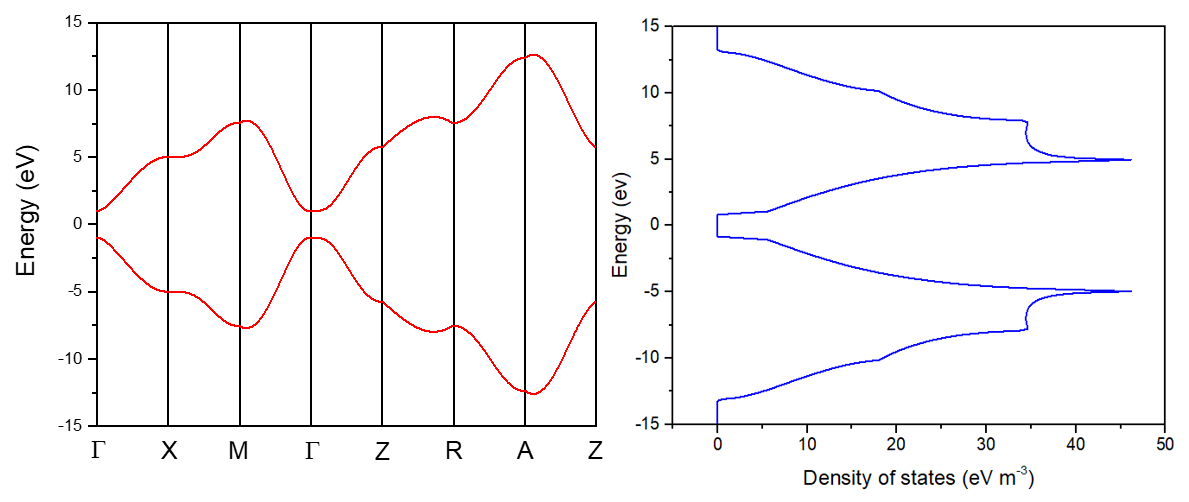}
    \caption{Energy bands through the high-symmetry points (left) and density of states (right) for the model in Eq. \ref{eq:TBmodel}.}
    \label{fig:DoS_and_bands}
\end{figure*}

To guide the search for a minimal tight binding model, we start by constraining the band edge model and then we convert it to a lattice model by the usual procedure. Assuming the conduction and valence band states transform as $A_1$ (scalar) irreps of $C_{6v}$, the allowed terms in the Hamiltonian are $v_{1,z}$, $v_{2,z}$ and $A_{\alpha,zz}$, $A_{\alpha,xx} = A_{\alpha,yy}$ for $\alpha = 1,2,3$. As the simplest choice we can take 
\begin{align}
\label{eq:A1model}
\vec f = (\alpha k_z + \beta (k^2_x+k_y^2),\alpha k_z + \beta(k^2_x+k_y^2) ,M + \gamma k^2_z).
\end{align}
We now consider a simple tight binding Hamiltonian whose $k=0$ expansion leads to the previous band edge model, which is simply
\begin{align}
f_1 &= t_1 \sin k_z a + t_2(2-\cos k_x a - \cos k_y a), \\
f_2 &= t_1 \sin k_z a + t_2(2-\cos k_x a - \cos k_y a), \\
f_3 &= M + t_3(1 - \cos k_z a), \label{eq:TBmodel}
\end{align}
where we assumed the same lattice constant $a$ for all three directions. This results in the parameters $v_{1,z} =v_{2,z}  = t_1 a$, $A_{1,xx} = A_{2,xx} = A_{1,yy} = A_{2,yy} =  t_2 a^2/2$ and $A_{3,zz} = t_3a^2/2$ and the rest vanish. From Eq. \ref{asy} we see that in this simple model the $k$ to $-k$ asymmetry occurs only in the $z$ direction and scales with the coefficient $t_1$. We take $t_1=1$, $t_2 = t_3 = 2$ and $M=1$. The band dispersion and density of states (DOS) of this model are shown in Fig. \ref{fig:DoS_and_bands}.

\begin{figure*}
    \centering
    \includegraphics[width=\textwidth]{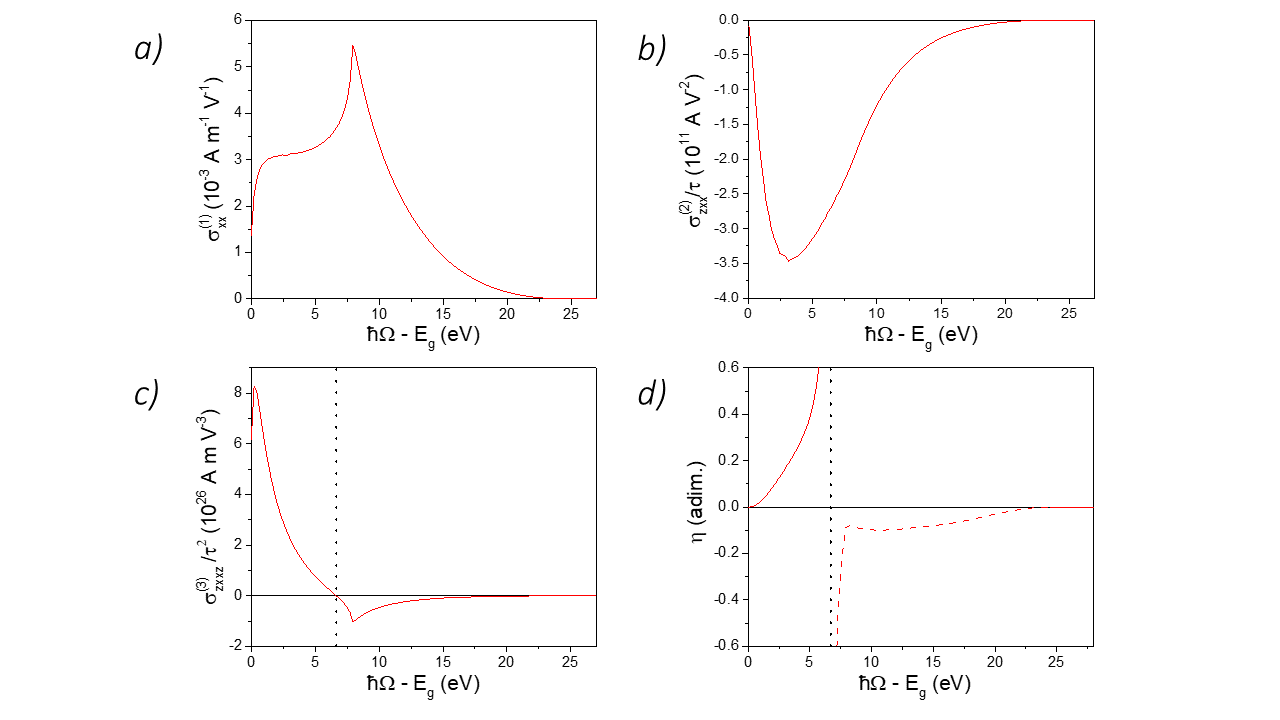}
    \caption{Linear conductivity (a),  photoconductivity (b),  jerk current (c) and power efficiency (d) for the full bands and light incident along y generating a current along x.  The black dotted lines indicates the position of the jerk current's zero which causes the efficiency to diverge at that point, and the red dotted line indicates the non-physical negative efficiency region .}
    \label{fig:full-bands}
\end{figure*}

This model can now be used to compute the different optical responses (absorption, magnetic injection and jerk current) given in Eqs. \ref{Eq:absorption}-\ref{Eq:jerk}, which are shown in Fig. \ref{fig:full-bands} $a)-c)$. Injection and jerk currents are plotted divided by the corresponding factor of $\tau$ to plot a $\tau$-independent quantity. The results show the typical non-monotonic behavior with cusps associated to Van Hove singularities. In addition, we observe that the jerk current has a sign change at $\approx 6.7$ eV, which is made possible by the fact that the second derivative of the energy difference $\partial_i\partial_j \epsilon_{nm}$ can change sign accross the Brilloin Zone. Finally, \ref{fig:full-bands}d shows the efficiency computed from these three quantities. Interestingly we observe the efficiency quickly raises to large values as a function of $\Omega$, eventually diverging where the jerk current vanishes. After that point, the efficiency is not well defined because the jerk current has opposite sign to the absorption, and it is plotted as a dashed line only for completeness. 

\section{Discussion}

In this work, we have obtained a simple expression for the energy harvesting efficiency of a material, which is an intrinsic band structure property. In the relaxation time approximation and at zero temperature, this requires a magnetic non-centrosymmetric material, where current generation is dominated by the injection mechanism. Energy harvesting from unpolarized light further requires a material in one of the polar point groups, $C_1,C_2,C_s,C_{2v},C_3,C_{3v},C_4,C_{4v},C_6,C_{6v}$. Insulating polar magnets, which have been studied in the context of magnetoelectric and multiferroic effects ~\cite{Eerenstein06} are therefore the class of materials relevant for this effect. 

It is worth noting that the symmetry that constrains the injection current tensor is only the non-magnetic part of the magnetic point group. Elements of the magnetic point group which are combined with time-reversal symmetry only constrain the circular components of the injection mechanism, but have no effect on the linear components considered in this work. For example, the magnetic semiconductor MnPS$_3$~\cite{Ressouche10} which has a gap of $\sim$ 3 eV, has magnetic point group $2'/m$. The only non-magnetic element of this group is a mirror plane $m$, so the current can flow in any direction parallel to the mirror plane.  

To estimate the photoconductivty, we have considered the known expression for the jerk current, which can be interpreted as the correction to the DC linear conductivity to first order in light intensity. We have seen that at finite light frequencies $\Omega$, the photoconductivity can be negative, which would formally imply that the photocurrent is not opposed by the generated voltage but rather further increased by it. This unphysical situation is produced because we have computed only linear conductivity (to first order in applied voltage), while in practice the conductivity will be non-linear at higher bias and eventually recover the positive sign, leading to a finite open circuit voltage. Attempting this calculation is beyond the scope of this work, which should therefore be understood to apply only for the frequency regions where the photoconductivity is positive. A related reason why negative photoconductivity is obtained is that we do not consider finite temperature effects and relaxation of the electron population. In that case, photocarriers excited into energies with negative photoconductivity would eventually relax to the bottom of the band, where the photoconductivity is again positive. Finally, it should be born in mind that the actual efficiency for energy harvesting from sunlight requries an integral weighted by the solar spectrum, and that as usual the total efficiency is maximized when the peak efficiency is aligned with the solar peak.   

Because of these reasons, our calculations should not be interpreted as a realistic prediction for material efficiencies, but rather as simple proof of principle that efficiencies can in fact be computed from band structure properties. This can help to identify band structure traits which lead to increased efficiency, and it can also allow for first principles implementations for material screening, which should help to identify general classes of materials with efficient photovoltaic responses. We hope that our work will stimulate more refined calculations and provide a path to explore the potential of magnetic injection photovoltaics. 

\vspace{1cm}

\textit{Acknowledgements}- We acknowledge Takahiro Morimoto and Julen Iba\~{n}ez-Azpiroz for useful discussions about our work. A.~J. C.-H. gratefully acknowledges support from the Donostia International Physics Centre (DIPC) internship programme where this work was initiated. A.~G.~G. acknowledges financial support by the ANR under the grant ANR-18-CE30-0001-01~(TOPODRIVE). F.J. acknowledges funding from the Spanish MCI/AEI/FEDER through grant PGC2018-101988-B-C21 and from Diputaci\'{o}n de Gipuzkoa through grant Gipuzkoa NEXT 2021-100-000070-01. 

\appendix

\section{General optical responses}\label{AppOptics}

The current response to time dependent electric field can be expressed in a power series in the field as
\begin{align}
J_i(t) = J_i^{(1)}(t) + J_i^{(2)}(t) +J_i^{(3)}(t) + \ldots
\end{align}
where
\begin{align}
J_i^{(1)}(t) &= \int \frac{d\omega_1}{2\pi}\sigma_{ij}^{(1)} E_j(\omega_1)e^{-i \omega_{1}t} , \\
J_i^{(2)}(t) &= \int \frac{d\omega_1}{2\pi}\frac{d\omega_2}{2\pi}\sigma_{ijk}^{(2)} E_j(\omega_1)E_k(\omega_2)e^{-i \omega_{12}t},\\
J_i^{(3)}(t) &= \int \frac{d\omega_1}{2\pi}\frac{d\omega_2}{2\pi}\frac{d\omega_3}{2\pi} \sigma_{ijkl}^{(3)} E_j(\omega_1)E_k(\omega_2)E_l(\omega_3)e^{-i \omega_{123}t},
\end{align}
where $\sigma^{(n)}_{ij \ldots}$ is the $n$-th order conductivity, $\omega_{12} = \omega_1 + \omega_2$ and $\omega_{123} = \omega_1 + \omega_2 + \omega_3$.

Optical absorption is measured by the real part of the optical conductivity $\sigma^{(1)}_{ij}$, which is given by 
\begin{equation}
\sigma_{ij}^{(1)}(\omega_1) = e^2 \pi \omega_1 \int \frac{d^3k}{(2\pi)^3} \sum_{n \neq m} f_{mn} {\rm Re}[r^i_{nm}r^j_{mn}] \delta(\hbar \omega_1 - \epsilon_{nm}).
\end{equation}
In the presence of absorption at frequency $\omega_1 = \Omega$, the intensity of the applied light beam decays as $I(z) = E(0) e^{-2\alpha z}$ with the attenuation constant \cite{deJuan17}
\begin{align}
\alpha = \frac{\Omega}{c} \sqrt{2(-{\rm Re}[\epsilon]+|\epsilon|)},
\end{align}
where the dielectric constant is related to the conductivity as $\epsilon = 1-\tfrac{ \sigma_{xx}^{(1)}}{i\Omega \epsilon_0}$. Assuming a negligible imaginary part of $\sigma_{ij}^{(1)}$ and in the high frequency limit we have
\begin{align}
\alpha =  \frac{\Omega}{c} \sqrt{2\left(-1+\sqrt{1+\left(\tfrac{ \sigma_{xx}^{(1)}}{\Omega \epsilon_0}\right)^2}\right)} \approx \frac{ \sigma_{xx}^{(1)}}{c \epsilon_0}.
\end{align}

The second order conductivity $\sigma^{(2)}_{ijk}$ determines the photocurrent in the limit $\omega_1 \rightarrow -\omega_2$. In this limit, the response tensor is dominated by a pole in $1/\omega_{12}$ as \cite{DFG20}

\begin{align}
&\sigma_{ijk}^{(2)}(\omega_1,\omega_2) = \frac{\pi}{-i\omega_{12}}  \dfrac{e^3}{\hbar^2} \frac{d^3k}{(2\pi)^3} \nonumber \\
&\sum_{n \neq m} f_{nm}\partial_a\epsilon_{mn} r^c_{nm}r^b_{mn}\delta(\hbar \omega_1 - \epsilon_{nm}),
\end{align}
which is known as the injection current. This tensor contains both a real, symmetric part, which leads to an LPGE, and an imaginary, antisymmetric part, which leads to a CPGE.  In the presence of dissipation, we account for a finite lifetime $\tau$ with the replacement $\omega_i \rightarrow \omega_i + i/\tau$, interpreted in terms of adiabatic switching \cite{Passos18}. It should be noted that this choice leads to $\omega_{12}\rightarrow \omega_{12} + 2i\tau$ and $\omega_{123} \rightarrow \omega_{123} + 3i\tau$, with numerical prefactors that are different from those obtained from equations of motion supplemented by a scattering term \cite{Passos18}. 

The contribution to the DC current from a monochromatic beam of frequency $\Omega$

\begin{align}
E_i(\omega) &= \int dt e^{i \omega t}(\tilde{E}_i e^{\Omega t}+ \tilde{E}_i^* e^{-i\Omega t}) \nonumber\\
&= 2\pi (\tilde{E}_i \delta(\omega+\Omega)+ \tilde{E}_i^* \delta(\omega-\Omega)),
\end{align}
is
\begin{align}
J_i^{(2)}(t) &= \sigma_{ijk}^{(2)}(\Omega,-\Omega) \tilde{E}_j \tilde{E}_k^*+  \sigma_{ijk}^{(2)}(-\Omega,\Omega) \tilde{E}_j^* \tilde{E}_k  \nonumber \\
&= \sigma_{ijk}^{(2)}(\Omega,-\Omega) (\tilde{E}_j \tilde{E}_k^* + \tilde{E}_k \tilde{E}_j^*),
\end{align}
where we have only considered the LPGE contribution, so that $\sigma_{ijk}^{(2)}(\Omega,-\Omega) = \sigma_{ikj}^{(2)}(-\Omega,\Omega)$, with
\begin{align}
&\sigma_{ijk}^{(2)}(\Omega,-\Omega) = \tau  \dfrac{ \pi e^3}{2\hbar^2} \int \frac{d^3k}{(2\pi)^3} \nonumber \\
&\sum_{n \neq m} f_{nm}\partial_a\epsilon_{mn} {\rm Re}[r^c_{nm}r^b_{mn}]\delta(\hbar \Omega - \epsilon_{nm}).
\end{align}

The third order conductivity $\sigma^{(3)}_{ijkl}$ is responsible for the photoinduced DC conductivity of a semiconductor \cite{Fregoso19,Ventura20} in the limit where one of the fields has zero frequency.  The leading divergence of the full three-frequency response is given by 
\begin{align}
&\sigma_{ijkl}^{(3)}(\omega_1,\omega_2,\omega_3) = -\frac{e^4\pi}{3\hbar^3} \frac{1}{\omega_{123}}\frac{1}{\omega_{12}}  \int \frac{d^3k}{(2\pi)^3} \nonumber\\
&\sum_{n\neq m} \partial_{i} \partial_l \epsilon_{nm} {\rm Re}[r^i_{nm}r^j_{mn}f_{mn}] \delta(\hbar \omega_1 - \epsilon_{nm}) . \label{a1}
\end{align}

Similarly as before, to obtain the second order response to light of polarization $\tilde{E}_i$ in the presence of a constant electric field $E^0_i$ we take
\begin{align}
&E_i(\omega) = \int dt e^{i \omega t}(E^0_i + \tilde{E}_i e^{\Omega t}+ \tilde{E}_i^* e^{-i\Omega t}) \nonumber\\
&= 2\pi (E^0_i \delta(\omega) + \tilde{E}_i \delta(\omega+\Omega)+ \tilde{E}_i^* \delta(\omega-\Omega)).
\end{align}
The Jerk current is the leading divergence of Eq.  \ref{a1} (the terms proportional to $\tau^2$) which is linear in $E^0_i$, and is given by the terms where $\omega_{12} = 0 + 2i/\tau$, $\omega_3 = 0 + i/\tau$,  

\begin{align}
J_i(t) &= \sigma_{ijkl}^{(3)}(\Omega,-\Omega,0) \tilde{E}_j \tilde{E}_k^* E^0_l +  \sigma_{ijkl}^{(3)}(-\Omega,\Omega,0) \tilde{E}_j^* \tilde{E}_k E^0_l \nonumber \\
&= \sigma_{ijkl}^{(3)}(\Omega,-\Omega,0) (\tilde{E}_j \tilde{E}_k^* + \tilde{E}_k \tilde{E}_j^*)  E^0_l ,
\end{align}
where we used that $\sigma_{ijkl}^{(3)}(\Omega,-\Omega,0) = \sigma_{ikjl}^{(3)}(-\Omega,\Omega,0)$. Evaluating the frequencies we get
\begin{align}
&\sigma_{ijkl}^{(3)}(\Omega,-\Omega,0) =\tau^2 \frac{\pi e^4}{18 \hbar^3} \int \frac{d^3k}{(2\pi)^3} \nonumber\\
&\sum_{n\neq m} \partial_{i} \partial_l \epsilon_{nm}{\rm Re}[r^i_{nm}r^j_{mn}f_{mn}] \delta(\hbar \Omega - \epsilon_{nm})  .
\end{align}

\section{Identities for optical matrix elements}\label{App:Identities}

In this Appendix we summarize well known identities for optical matrix elements we used in this work. From the first derivative of Hamiltonian matrix elements 
\begin{equation}
\partial_{k_a} \left<n|H|m\right> = \delta_{nm} \partial_{k_a}\epsilon_n ,
\end{equation}
one obtains that 
\begin{align}
v^a_{nn} &=\partial_{k_a}\epsilon_n ,\\
v^a_{nm} & = i r^a_{nm}\epsilon_{nm} .
\end{align}
The last equation allows one to trade $r_{nm}^i$ which depends on wavefunction derivatives, for $v_{nm}^i$, which only depends on Hamiltonian derivatives. From the second derivative of Hamiltonian matrix elements
\begin{equation}
\partial_{k_b}\partial_{k_a} \left<n|H|m\right> = \delta_{nm} \partial_{k_b}\partial_{k_a}\epsilon_n ,
\end{equation}
we obtain the tight binding extension to the effective mass theorem
\begin{equation}
\partial_{k_b} v^a_{nn} = v_{nn}^{ba} - \sum_{p\neq n} \frac{v^a_{np} v^b_{pn}+v_{np}^bv^a_{pn}}{\epsilon_{pn}} = \partial_{k_b}\partial_{k_a}\epsilon_n ,
\end{equation}
with $v_{nm}^{ba}=\left<n|\partial_{k_b}\partial_{k_a}H|m\right>$. This identity is particularly useful for two band models where one can extract the symmetric product of velocity matrix elements found in the response function formulas 
\begin{equation}
r^a_{12} r^b_{21}+r_{12}^b r^a_{21} = \frac{v^a_{12} v^b_{21}+v_{12}^bv^a_{21}}{\epsilon_{21}^2} = - \frac{v_{11}^{ba}-\partial_{k_b}\partial_{k_a}\epsilon_1}{ \epsilon_{12}}.\label{diag}
\end{equation}

With these identities, the expressions for the photocurrent tensors in Appendix \ref{AppOptics} can then be computed. 

\section{General form of response functions for a two-band model}\label{App:TwoBand}

We consider the two band model in the main text. Applying the identities discussed in Appendix \ref{App:Identities}, the expressions for the optical responses for two bands reduce to 

\begin{widetext}
\begin{gather}
    \sigma_{ij} (\Omega) = - e^2 \pi \Omega \int_k \frac{v_{11}^{ij}-\partial_i \partial_j \epsilon_1}{\epsilon_{12}} \delta(\hbar\Omega-\epsilon_{12})\frac{d^3k}{(2\pi)^3}, \\
    \sigma_{ijk} (\Omega) = - \tau \dfrac{ \pi e^3}{2\hbar^2} \int_k \partial_i(\epsilon_{12}) \frac{v_{11}^{jl}-\partial_j \partial_l \epsilon_1}{\epsilon_{12}} \delta(\hbar\Omega-\epsilon_{12})\frac{d^3k}{(2\pi)^3},\\
    \sigma_{ijkl} (\Omega) = - \tau^2 \frac{\pi e^4}{18 \hbar^3} \int_k \partial_i\partial_p(\epsilon_{12}) \frac{v_{11}^{jl}-\partial_j \partial_l \epsilon_1}{\epsilon_{12}} \delta(\hbar\Omega-\epsilon_{12})\frac{d^3k}{(2\pi)^3}.
\end{gather}
\end{widetext}

We now seek to fully express these integrals in terms of the Hamiltonian components $f_{\alpha}$. To start, the wave functions of the two-band $H$ considered are
\begin{align}
\psi_{n} =  \frac{1}{\sqrt{2\epsilon}}(-\eta \sqrt{\epsilon -\eta f_z} \:, e^{i \phi_{k} }\sqrt{\epsilon +\eta f_z}),
\end{align}

with energies $\epsilon_1 = f_0 + \epsilon$ and $\epsilon_2 = -\epsilon_1$ and with $n=1,2$, $\eta=(-1)^n$, and $\phi_{k} = \arctan (f_y/f_x)$. From this we can write 
\begin{align}
\left<n|\sigma_\alpha|n\right> &= - \eta \frac{f_\alpha}{\epsilon_1}, 
\end{align}
so for $\epsilon_1$ we have
\begin{align}
\left<1|\sigma_\alpha|1\right> &= \frac{f_\alpha}{\epsilon}. 
\end{align}
This allows us to compute

\begin{align}
v_{11}^{ab} = \sum_\alpha (\partial_a \partial_b  f_\alpha) \left<1|\sigma_\alpha|1\right> = \sum_\alpha (\partial_a \partial_b  f_\alpha )\frac{f_\alpha}{\epsilon} .\label{eq:v-in-fs}
\end{align}

Next we also use that for $f_0=0$

\begin{align}
\partial_a \partial_b \epsilon_1 = \frac{\partial_a f_\alpha \partial_b f_\alpha + f_\alpha \partial_a\partial_b f_\alpha}{\epsilon} - \frac{ f_\alpha f_\beta \partial_a f_\alpha  \partial_b f_\beta}{\epsilon^3}, \label{eq:energy-in-fs}
\end{align}
so
\begin{equation}
\frac{v^a_{12} v^b_{21}+v_{12}^bv^a_{21}}{\epsilon_{21}^2} = \frac{   \partial_a f_\alpha \partial_b f_\alpha}{2\epsilon^2}  - \frac{f_\alpha \partial_a f_\alpha f_\beta \partial_b f_\beta}{2\epsilon^4}
\end{equation}
These expressions can then be used to compute the responses of any two band model. 

\section{Scaling near the band edge}
\label{App:Scaling}

The generic band edge model can be used to compute the dominant contribution to all optical responses as the frequency approaches the band edge, i.e. defining $\delta \Omega = \Omega - E_g$, we compute Eqs. \ref{Eq:absorption}-\ref{Eq:jerk} in the $\delta \Omega \ll 1$ limit. To do so, we first expand the integrand to lowest order in $k$
\begin{align}
\epsilon &= M + \frac{1}{2M}(A_{ij}k_ik_j + B_{ijk}k_ik_jk_k) + O(k^4), \\
&{\rm Re}[r^i_{12}r^j_{12}] = \frac{1}{4M^2}(C_{ij}+D_{ijk}k_k) + O(k^2),
\end{align}
where 
\begin{align}
A_{ij} &= v_{1,i} v_{1,j} + v_{2,i} v_{2,j} +2MA_{3,ij},  \\
B_{ijk} &= 2A_{1,ij}v_{1,k}+2A_{2,ij}v_{2,k}, \\
C_{ij} &= v_{1,i} v_{1,j} + v_{2,i} v_{2,j} ,\\
D_{ijk} &= 2(v_{1,i}A_{1,jk} + v_{1,j}A_{1,ik} + v_{2,i}A_{2,jk} + v_{2,j}A_{2,ik}),
\end{align}
and the lowest order terms in the integrand of Eqs. \ref{Eq:absorption}-\ref{Eq:jerk} are 
\begin{align}
    \alpha &\propto  \Omega \int \frac{d^3k}{(2\pi)^3} \frac{C_{ij}}{4M^2} \delta(\hbar \Omega - \epsilon_{12}) ,\\
\kappa_i & \propto \int \frac{d^3k}{(2\pi)^3} \left[ \frac{A_{im}C_{jk} k_m}{4M^3}\right. \nonumber \\&\left.+\frac{(2A_{im}D_{jkl} + 3B_{ilm} C_{jk})k_mk_l}{8M^3}\right]\delta(\hbar \Omega - \epsilon_{12}),\label{eq:injlow}\\
\Sigma_{ij} & \propto \int \frac{d^3k}{(2\pi)^3}\frac{ A_{il}C_{jk}}{2M^3}\delta(\hbar \Omega - \epsilon_{12}),
\end{align}
where the need to keep quadratic terms in $\kappa_i$ will become clear shortly. To evaluate the $\delta$-function in the integrals we use
\begin{equation}
    \delta (g(x)) = \sum_{x_i} \frac{\delta(x - x_i)}{\vert \partial_{x} g(x_i) \vert},
\end{equation}
where $x_i$ are the zeros of $g(x)$. In our case, the zeros in $k$ are obtained from $\Omega -\epsilon_{12} =0$, which to leading order in $\delta \Omega$ is 
\begin{align}
\delta \Omega M&= A_{ij}k_ik_j + B_{ijk}k_ik_jk_k.
\end{align}
To solve this we go to polar coordinates and perform a coordinate transformation $k'_i = U_{ij}k_j$ that diagonalizes $X_{ij}$ so that $k_i X_{ij} k_j = k'^2$ and 
\begin{align}
\delta \Omega M&= k'^2 + Y_{ijk}(U^{-1}k')_i(U^{-1}k')_j(U^{-1}k')_k \nonumber \\
&=  k'^2 + f(\theta',\phi')k'^3.
\end{align}
This has a simple perturbative solution
\begin{align}
k'(\delta \Omega) = (\delta \Omega M)^{1/2} - \frac{ \delta \Omega M}{2}f(\theta',\phi').\label{pert}
\end{align}
For the integrals in Eq. \ref{Eq:absorption} and Eq. \ref{Eq:jerk}, the last term is not needed as $k'(\delta \Omega) = (\delta \Omega M)^{1/2}$ already provides the leading contribution. Evaluating the $\delta$ function, we find both integrals scale as
\begin{align}
    \alpha   \propto \delta \Omega^{1/2}, \\
    \Sigma_{ij} \propto \delta \Omega^{1/2}.
\end{align}
The leading term for the magnetic injection has two contributions. First, the linear term in momentum in Eq. \ref{eq:injlow} can integrate to something finite when evaluating the $\delta$-function due to the second term in Eq. \ref{pert}, which gives rise to a term scaling as $(\delta \Omega)^{3/2}$. Second, the quadratic term in momentum in Eq. \ref{eq:injlow} also leads to another term scaling as $(\delta \Omega)^{3/2}$, so overall
\begin{align}
    \kappa_i   \propto \delta \Omega^{3/2} .
\end{align}
The analytic expressions for these terms are cumbersome to show, but the conclusion is that the efficiency near the band edge is proven to scale as 
\begin{equation}
    \eta \propto  \frac{(\delta \Omega)^3}{(\delta \Omega)^{1/2}(\delta \Omega)^{1/2}} = (\delta \Omega)^2,
\end{equation}
and therefore vanishes at the band edge.

\bibliography{weyl}

\newpage

\end{document}